\pdfoutput=1
\RequirePackage{ifpdf}
\ifpdf %We are running pdfTeX in pdf mode
\documentclass[pdftex]{sigma}
\else
\documentclass{sigma}
\fi

\DeclareMathOperator{\diag}{diag}

\numberwithin{equation}{section}

\begin{document}

\newcommand{\arXivNumber}{1402.0397}

\allowdisplaybreaks

\renewcommand{\thefootnote}{$\star$}

\renewcommand{\PaperNumber}{106}

\FirstPageHeading

\ShortArticleName{$\kappa$-Deformed Phase Space, Hopf Algebroid and Twisting}

\ArticleName{$\boldsymbol{\kappa}$-Deformed Phase Space, Hopf Algebroid\\
and Twisting\footnote{This paper is a~contribution to the Special Issue on Deformations of Space-Time and its
Symmetries.
The full collection is available at \href{http://www.emis.de/journals/SIGMA/space-time.html}
{http://www.emis.de/journals/SIGMA/space-time.html}}}

\Author{Tajron JURI\'C~$^\dag$, Domagoj KOVA\v{C}EVI\'C~$^\ddag$ and Stjepan MELJANAC~$^\dag$}

\AuthorNameForHeading{T.~Juri\'c, D.~Kova\v{c}evi\'c and S.~Meljanac}

\Address{$^\dag$~Rudjer Bo\v{s}kovi\'c Institute, Bijeni\v cka cesta 54, HR-10000 Zagreb, Croatia}
\EmailD{\href{mailto:tajron.juric@irb.hr}{tajron.juric@irb.hr}, \href{mailto:meljanac@irb.hr}{meljanac@irb.hr}}

\Address{$^\ddag$~University of Zagreb, Faculty of Electrical Engineering and Computing,\\
\hphantom{$^\ddag$}~Unska 3, HR-10000 Zagreb, Croatia}
\EmailD{\href{mailto:domagoj.kovacevic@fer.hr}{domagoj.kovacevic@fer.hr}}

\ArticleDates{Received February 21, 2014, in f\/inal form November 11, 2014; Published online November 18, 2014}

\Abstract{Hopf algebroid structures on the Weyl algebra (phase space) are presented.
We def\/ine the coproduct for the Weyl generators from Leibniz rule.
The codomain of the coproduct is modif\/ied in order to obtain an algebra structure.
We use the dual base to construct the target map and antipode.
The notion of twist is analyzed for~$\kappa$-deformed phase space in Hopf algebroid setting.
It is outlined how the twist in the Hopf algebroid setting reproduces the full Hopf algebra structure
of~$\kappa$-Poincar\'e algebra.
Several examples of realizations are worked out in details.}


\Classification{81R60; 17B37; 81R50}

\renewcommand{\thefootnote}{\arabic{footnote}}
\setcounter{footnote}{0}

\vspace{-2mm}

\section{Introduction}

\looseness=-1
Motivation for studying noncommutative (NC) spaces is related to the fact that general theory of relativity together
with Heisenberg uncertainty principle leads to the uncertainty of position coordinates
itself $\bigtriangleup x_{\mu} \bigtriangleup x_{\nu} > l^2_{\text{Planck}}$~\cite{Doplicher, Doplicher1}.
This uncertainty in the position can be reali\-zed via NC coordinates.
There are also arguments based on quantum gravity~\cite{Doplicher, Doplicher1, kempf}, and string theory
models~\cite{boer, Witten}, which suggest that the spacetime at the Planck length is quantum, i.e.~noncommutative.

We will consider a~particular example of NC space, the so called~$\kappa$-Minkowski
spacetime~\cite{bp09,byk2009,klry08,km11,kmps12,Lukierski-2,Lukierski-3,Lukierski-1,Majid-Ruegg,mks07,mssg08,mss12,Meljanac-4},
which is a~Lie algebraic deformation of the usual Minkowski spacetime.
Here,~$\kappa$ is the deformation parameter usually interpreted as Planck mass or the quantum gravity scale.
Investigations of physical theories on~$\kappa$-Minkowski spacetime leads to many new properties, such as: modif\/ication
of particle statistics~\cite{kappaSt, k2, gghmm0809,Gumesa, k4, k3}, deformed electrodynamics~\cite{h,hjm11}, NC quantum
mechanics~\cite{andrade, andrade1, harisiva, kuprijanov1,kuprijanov}, and quantum gravity ef\/fects~\cite{bgmp10,dolan,BTZ,
hajume, solo}. $\kappa$-Minkowski spacetime is also related to doubly-special and deformed relativity
theories~\cite{ Amelino-Camelia-1, Amelino-Camelia-2, bojowald, Kowalski-Glikman-2,Kowalski-Glikman-1}.

The symmetries of~$\kappa$-Minkowski spacetime are described via Hopf algebra setting and they are encoded in
the~$\kappa$-Poincar\'e--Hopf algebra
(in the same sense as are the symmetries of Minkowski spacetime encoded in the Poincar\'e--Hopf algebra).
A~Hopf algebra is a~bialgebra equipped with an antipode map satisfying the Hopf axiom.
The bialgebra is an (unital, associative) algebra which is also a~(conunital, coassociative) coalgebra such that certain
compatibility conditions are satisf\/ied.
The antipode is an antihomomorphism of the algebra structure (an antialgebra homomorphism).
Hopf algebras are used in various areas of mathematics and physics for f\/ifty years.
See~\cite{b08, Majid-1} for some examples.

It turns out that the notion of the Hopf algebra is too restrictive and it has to be generalized.
For example, it is shown that the Weyl algebra (quantum phase space) can not have a~structure of a~Hopf algebra.
Namely, the whole phase space (Weyl algebra) generated by $p_{\mu}$ and $x_{\mu}$ (or~$\hat{x}_{\mu}$) can not be
equipped with the Hopf algebra structure, since one can not include $\bigtriangleup x_{\mu}$ in a~satisfactory way, i.e.~the
notion of Hopf algebra is too restrictive for the whole phase space (Weyl algebra).
Several types of generalizations are possible: quasi-Hopf algebras, multiplier Hopf algebras and weak Hopf algebras.
Our construction is very similar to the structure of the Hopf algebroid def\/ined by Lu in~\cite{l95}.

\looseness=-1
Lu was inspired by the notion of the Poisson algebroid from the Poisson geometry.
Namely, some Hopf algebras are quantization of the Poisson groups.
Now, Hopf algebroids can be considered as the quantization of the Poisson groupoids.
Lu introduces two algebras: the base algebra~$A$ and the total algebra~$H$.
One can consider the total algebra~$H$ as the algebra over the base algebra~$A$.
The left and right multiplications are given by the source and the target maps.
Hence, the coproduct $\bigtriangleup$ is def\/ined on the total algebra~$H$ and the image lies in
$H\otimes_AH$ which is an $(A,A)$-bimodule but not an algebra.
Namely, $H\otimes_AH$ is the quotient of $H\otimes H$ by the right ideal.
G.~B\"{o}hm and K.~Szlach\'{a}nyi in~\cite{bs04} considered the same structure as Lu did, but they changed the def\/inition of the antipode.
For more comprehensive approach, see~\cite{b08}.
Let us mention that some ideas existed before the def\/inition of Lu in which the base algebra or both the base algebra
and the total algebra had to be commutative (see~\cite{b08, l95} and references therein).
Bialgebroid is equivalent to the notion of $\times_{A}$-bialgebra introduced much earlier by Takeuchi in~\cite{takeuci}.

One can analyze the structure of the Hopf algebra by twists.
See~\cite{admw051, admw05} for more details.
P.~Xu in~\cite{x01} applies the twist to the bialgebroid (which he calls Hopf algebroid although he does not have the
antipode).
It is important to mention that Xu uses the def\/inition of the bialgebroid which is equivalent to the def\/inition
from~\cite{l95}.

In~\cite{km11}~$\kappa$-Minkowski spacetime and Lorentz algebra are unif\/ied in a~unique Lie algebra.
Realizations and star products are def\/ined and analyzed in general and specially, their relation to coproduct of the
momenta is pointed out.

The deformation of Heisenberg algebra and the corresponding coalgebra by twist is performed in~\cite{mss12}.
Here, the so called tensor exchange identities are introduced and coalgebras for the generalized Poincar\'e algebras are constructed.
The exact universal~$R$-matrix for the deformed Heisenberg (co)algebra is found.

The quantum phase space (Weyl algebra) and its Hopf algebroid structure is analyzed in~\cite{jms13b}.
Unif\/ication of~$\kappa$-Poincar\'e algebra and~$\kappa$-Minkowski spacetime is done via embedding into quantum phase space.
The construction of~$\kappa$-Poincar\'e--Hopf algebra and~$\kappa$-Minkowski spacetime using Abelian twist in the Hopf
algebroid approach has been elaborated.

Twists, realizations and Hopf algebroid structure of~$\kappa$-deformed phase space are discussed in~\cite{jms13a}.
It is shown that starting from a~given deformed coalgebra of commuting coordinates and momenta one can construct the
corresponding twist operator.

In the present paper, the total algebra is the Weyl algebra $\hat{\mathcal{H}}$ and the base algebra is the
subalgebra $\hat{\mathcal A}$ generated by noncommutative coordinates $\hat{x}_\mu$.
The construction of the target map is obtained via dual realizations.
The codomain of the coproduct is changed.
We take a~quotient of the image of the coproduct instead of quotient of
$\hat{\mathcal{H}}\otimes \hat{\mathcal{H}}$.
As a~consequence, the right ideal by which Lu~\cite{l95} has taken the quotient is now two-sided and the codomain of the
coproduct has the algebra structure.
The notion of the counit is related to realizations.
Furthermore, we manage to incorporate the twist in our construction, obtaining the Hopf algebroid structure from the twist.

This paper is structured as follows.
In Section~\ref{dps} we introduce the~$\kappa$-Minkowski spacetime and~$\kappa$-deformed phase space, and we establish the
connection between Leibniz rule and coproduct for the Weyl generators.
Also, the dual basis is introduced and elaborated.
The Hopf algebroid structure of~$\kappa$-deformed phase space $\hat{\mathcal{H}}$ and undeformed phase space
$\mathcal{H}$ is presented in Section~\ref{ha}.
In Section~\ref{tw} we f\/irst discuss the realizations and then we provide the twist operator in the Hopf algebroid approach.
It is shown that the twisted Hopf algebroid structure of phase space $\mathcal{H}$ is isomorphic to the Hopf algebroid
structure of $\hat{\mathcal{H}}$.
Finally, in Section~\ref{Section5}
we consider the~$\kappa$-Poincar\'e--Hopf algebra in the \textit{natural} realization (classical basis).
It is outlined how the twist in Hopf algebroid setting reproduces the full Hopf algebra structure of~$\kappa$-Poincar\'e algebra.
Also, we discuss the existence and properties of twist in all types of deformations (space-, time- and light-like).

\section[$\kappa$-deformed phase space]{$\boldsymbol{\kappa}$-deformed phase space}
\label{dps}

\subsection[$\kappa$-Minkowski spacetime]{$\boldsymbol{\kappa}$-Minkowski spacetime}
%\label{ms}

Let us denote coordinates of the $\kappa$-Minkowski spacetime by $\hat{x}_\mu$.
Latin indices will be used for the set $\{1,\ldots,n-1\}$ and Greek indices will be used for the set $\{0,\ldots,n-1\}$.
The Lorentz signature of the $\kappa$-Minkowski spacetime is def\/ined~by
$[\eta_{\mu\nu}]=\diag(-1,1,\ldots,1)$.
Let ${\mathfrak g}_{\kappa}$ be the Lie algebra generated by $\hat{x}_\mu$ such that
\begin{gather}
\label{b1}
[\hat{x}_\mu,\hat{x}_\nu]=i\left(a_\mu\hat{x}_\nu-a_\nu\hat{x}_\mu\right),
\end{gather}
where $a\in {\mathbb M}^n$.
The relation to $\kappa$\ mass parameter is $a_\mu=\frac1{\kappa} u_\mu$,
$u_\mu\in {\mathbb M}^n$ ($u^2=-1$ time-like, $u^2=1$ space-like and $u^2=0$ light-like).
The enveloping algebra $ {\mathcal U}( {\mathfrak g}_{\kappa})$ of ${\mathfrak
g}_{\kappa}$ will be denoted by~${\hat{\mathcal A}}$.

\subsection{Phase space}
\label{ps}

The momentum space ${\mathcal T}={\mathbb C}[[p_\mu]]$ is the commutative space generated~by
$p_\mu$ such that
\begin{gather}
\label{b4}
[p_\mu,\hat{x}_\nu]=-i\varphi_{\mu\nu}(p)
\end{gather}
is satisf\/ied for some set of real functions $\varphi_{\mu\nu}$ (see~\cite{jms13b,jms13a, km11} for details).
Let us recall that $\lim\limits_{a\rightarrow0}\varphi_{\mu\nu}=\eta_{\mu\nu}1$ and $\det \varphi\neq0$.
We also require that generators $\hat{x}_\mu$ and $p_\mu$ satisfy Jacobi identities.
This gives the set of restrictions on functions $\varphi_{\mu\nu}$ (see equation~(11) in~\cite{km11} or equation~(4) in~\cite{jms13a}).
The existence of such space ${\mathcal T}$ is analyzed in several papers~\cite{km11, mks07}.
One particularly interesting solution is the set $\{p^{\rm L}_\mu\}$ which is related to the so called \textit{left covariant}
realization~\cite{km11, mks07} where $\varphi_{\mu\nu}=\eta_{\mu\nu}Z^{-1}$, i.e.~\eqref{b4} leads to
\begin{gather}
\label{b5}
\big[p^{\rm L}_\mu,\hat{x}_\nu\big]=-i\eta_{\mu\nu}Z^{-1}.
\end{gather}
Here~$Z$ denotes the shift operator def\/ined~by
\begin{gather*}
[Z,\hat{x}_\mu]=ia_\mu Z,
\qquad
[Z,p_\mu]=0,
\end{gather*}
and for the \textit{left covariant} realization is given~by
\begin{gather*}
%\label{b6}
Z^{-1}=1+\big(a p^{\rm L}\big),
\end{gather*}
where we used $\big(a p^{\rm L}\big)\equiv a^\alpha p_\alpha^{\rm L}$.
The phase space ${\hat{\mathcal{H}}}$ is generated as an algebra by ${\hat{\mathcal A}}$ and
${\mathcal T}$ such that~\eqref{b1} and~\eqref{b4} are satisf\/ied.

Let $\blacktriangleright$ be the unique action of $\hat{\mathcal{H}}$ on $\hat{\mathcal{A}}$, such that
$\hat{\mathcal{A}}$ acts on itself by left multiplication and $t \blacktriangleright \hat{f} = [t, \hat{f}]
\blacktriangleright 1$ for all $t\in\mathcal{T}$ and $\hat{f}\in\hat{\mathcal{A}}$.
${\hat{\mathcal A}}$ can be considered as an ${\hat{\mathcal{H}}}$-module.

\subsection{Leibniz rule}
%\label{lr}

We have already mentioned that ${\hat{\mathcal{H}}}$ does not have the structure of the Hopf algebra, but it
is possible to construct the structure of the Hopf algebroid.
In this subsection we do the preparation for the coproduct which will be completely def\/ined in Section~\ref{ha}.
The formula for the coproduct can be built from the action ${\blacktriangleright}$ and the Leibniz rule (see~\cite[Section 2.3]{km11} and~\cite{jms13a}).
In~$\kappa$-Poincar\'e--Hopf algebra $\mathcal{U}_{\kappa}(\mathcal{P})$ (where $\mathcal{P}$ is generated by momenta
$p_{\mu}$ and Lorentz generators $M_{\mu\nu}$) the coproducts of momenta and Lorentz generators are unique and
$\bigtriangleup|_{\mathcal{U}_{\kappa}(\mathcal{P})}:
\mathcal{U}_{\kappa}(\mathcal{P})\rightarrow\mathcal{U}_{\kappa}(\mathcal{P})\otimes\mathcal{U}_{\kappa}(\mathcal{P})$.
However in the Hopf algebroid structure the coproduct of genera\-tors~$p_{\mu}$ and $\hat{x}_{\mu}$ are not unique, modulo
the right ideal ${\hat{\mathfrak{K}}}$ in~\eqref{JH}.

Let $\bigtriangleup(\hat{h})=\hat{h}_{(1)}\otimes \hat{h}_{(2)}$ for
$\hat{h}_{(1)},\hat{h}_{(2)}\in {\hat{\mathcal{H}}}$ (using Sweedler notation).
Then
\begin{gather}
\label{b7}
\hat{h} {\blacktriangleright}\big(\hat{f}\hat{g}\big)
=m\big({\bigtriangleup}\big(\hat{h}\big) {\blacktriangleright}\big(\hat{f}\otimes\hat{g}\big)\big)
=\big(\hat{h}_{(1)} {\blacktriangleright}\hat{f}\big)\big(\hat{h}_{(2)} {\blacktriangleright}\hat{g}\big)
\end{gather}
for $\hat{f},\hat{g}\in {\hat{\mathcal A}}$.

Now we recall the formula for the coproduct of $p_{\mu}$ def\/ined~by
$\bigtriangleup|_{\mathcal{T}}:\mathcal{T}\rightarrow\mathcal{T}\otimes\mathcal{T}$.
Then
\begin{gather}
\label{b8}
p_\mu {\blacktriangleright}\big(\hat{f}\hat{g}\big)=\big[p_\mu,\hat{f}\hat{g}\big] {\blacktriangleright}1=
\big(\big[p_\mu,\hat{f}\big]\hat{g}+\hat{f}[p_\mu,\hat{g}]\big) {\blacktriangleright}1=
\big[p_\mu,\hat{f}\big] {\blacktriangleright}\hat{g}+\hat{f}p_\mu {\blacktriangleright}\hat{g}.
\end{gather}
For example let us write the coproduct of $p_\mu^{\rm L}$.
One f\/inds by induction, starting with~\eqref{b5} that $\big[p_\mu^{\rm L}, \hat{f}\big]=\big(p_\mu^{\rm L}\blacktriangleright\hat{f}\big)Z^{-1}$,
$\forall\,\hat{f}\in\hat{\mathcal{A}}$.
Inserting this result in the r.h.s.\ of~\eqref{b8} and comparison with r.h.s.\
of~\eqref{b7} for $\hat{h}=p^{\rm L}_{\mu}$ gives
\begin{gather*}
\bigtriangleup\big(p^{\rm L}_\mu\big)=p^{\rm L}_\mu\otimes Z^{-1}+1\otimes p^{\rm L}_\mu.
\end{gather*}

Now, let us consider elements $\hat{x}_\mu$.
It is clear that
\begin{gather}
\label{b9}
\bigtriangleup(\hat{x}_\mu)=\hat{x}_\mu\otimes1
\end{gather}
since $\hat{x}_\mu {\blacktriangleright} (\hat{f}\hat{g} )= (\hat{x}_\mu\hat{f} )\hat{g}$.
Formula (33) from~\cite{km11} shows that
\begin{gather*}
\hat{x}_\mu {\blacktriangleright}\big(\hat{f}\hat{g}\big)
=\big(Z^{-1} {\blacktriangleright}\hat{f}\big)(\hat{x}_\mu {\blacktriangleright}
\hat{g})-a_\mu\big(p_\alpha^{\rm L} {\blacktriangleright}\hat{f}\big)(\hat{x}_\alpha {\blacktriangleright}\hat{g})
\end{gather*}
and\footnote{In Section~\ref{dha}, the coproduct will be def\/ined and~\eqref{b9} and~\eqref{b10} will be equal, since
both choices of coproducts of $\hat x_\mu$ belong to the same congruence class.}
\begin{gather}
\label{b10}
\bigtriangleup'(\hat{x}_\mu)= Z^{-1}\otimes\hat{x}_\mu-a_\mu p^{\rm L}_\alpha\otimes\hat{x}^\alpha.
\end{gather}

It is convenient to write~\eqref{b10} in the form
\begin{gather*}
\bigtriangleup'(\hat{x}_\mu)=O_{\mu\alpha}\otimes\hat{x}^\alpha,
\end{gather*}
where
\begin{gather}
\label{b11}
O_{\mu\alpha}=Z^{-1}\eta_{\mu\alpha}-a_\mu p^{\rm L}_\alpha.
\end{gather}
Hence, elements
\begin{gather}
\label{rmu}
{\hat{R}_\mu}=\hat{x}_\mu\otimes1-O_{\mu\alpha}\otimes\hat{x}^\alpha
\end{gather}
satisfy $m({\hat{R}_\mu}{\blacktriangleright}(\hat{f}\otimes\hat{g}))=0$ for all
$\hat{f},\hat{g}\in{\hat{\mathcal A}}$ where~$m$ denotes the multiplication ($m(\hat{f}
\otimes\hat{g})=\hat{f}\hat{g}$) and $(a\otimes b){\blacktriangleright}(\hat{f}\otimes\hat{g})=
(a{\blacktriangleright}\hat{f})\otimes(b{\blacktriangleright}\hat{g})$.
Then
\begin{gather}
\label{JH}
{\hat{\mathfrak{K}}}=\mathcal{U}_{+}(\hat{R}_{\mu}){\hat{\mathcal{H}}}\otimes{\hat{\mathcal{H}}}
\end{gather}
is the right ideal in ${\hat{\mathcal{H}}}\otimes{\hat{\mathcal{H}}}$.
Here we used that $\mathcal{U}_{+}({\hat{R}_\mu})$ is the universal enveloping algebra generated~by
${\hat{R}_\mu}$ but without the unit element.

It is important to emphasize that such derived coproduct is an algebra homomorphism
\begin{gather*}
\bigtriangleup\big(\hat{h}_1\hat{h}_2\big)=\bigtriangleup\big(\hat{h}_1\big)\bigtriangleup\big(\hat{h}_2\big)
\end{gather*}
for any $\hat{h}_1,\hat{h}_2\in{\hat{\mathcal{H}}}$ which enables us to def\/ine the formula for the coproduct for all elements
of~${\hat{\mathcal{H}}}$.

\subsection{Dual basis}
%\label{db}

In~\cite{km11} we have introduced the notion of the dual basis.
Let us recall some basic facts since it will be used for the def\/inition of the target map.
We def\/ine elements
\begin{gather}
\label{b12}
\hat{y}_\mu=\hat{x}^\alpha O^{-1}_{\mu\alpha},
\end{gather}
where
\begin{gather}
\label{b13}
O^{-1}_{\mu\alpha}=\big(\eta_{\mu\alpha}+a_\mu p^{\rm L}_\alpha\big)Z
\end{gather}
(it would be more precise to write $(O^{-1})_{\mu\alpha}$).
They have some interesting properties.
Since
\begin{gather}
\label{b16}
\hat{x}_\mu=\hat{y}^\alpha O_{\mu\alpha},
\end{gather}
$\hat{y}_\mu$ and $p_\mu$ form a~basis of ${\hat{\mathcal{H}}}$ (it would be more correct to say that power
series in $\hat{y}_\mu$ and $p_\mu$ form a~basis of ${\hat{\mathcal{H}}}$).
Elements $\hat{y}_\mu$ satisfy commutation relations similar to~\eqref{b1}:
\begin{gather*}
[\hat{y}_\mu,\hat{y}_\nu]=-i(a_\mu\hat{y}_\nu-a_\nu\hat{y}_\mu).
\end{gather*}
We call this basis the dual basis.

It is easy to check that $\hat{x}_\mu$ and $\hat{y}_\nu$ commute, i.e.\
\begin{gather}
\label{b17}
[\hat{x}_\mu,\hat{y}_\nu]=0.
\end{gather}
Also, the straightforward calculation shows that $O_{\mu\nu}$ and $O_{\lambda\rho}$ commute.
It remains to consider commutation relations among $O_{\mu\nu}$, $\hat{x}_\mu$ and $\hat{y}_\mu$.
The def\/inition of $O_{\mu\nu}$ yields $[O_{\mu\nu},\hat{x}_\lambda]=i(a_\mu\eta_{\lambda\nu}-a_\lambda
\eta_{\mu\nu})Z^{-1}=i(a_\mu O_{\lambda\nu}-a_\lambda O_{\mu\nu})$ and it shows that
\begin{gather}
\label{b18}
[O_{\mu\nu},\hat{x}_\lambda]=iC_{\mu\lambda}^{\ \ \alpha}O_{\alpha\nu},
\end{gather}
where $C_{\mu\lambda\alpha}=a_\mu\eta_{\lambda\alpha}-a_\lambda\eta_{\mu\alpha}$ stands for structure constants.
One can easily obtain
\begin{gather*}
\big[O_{\mu\nu}^{-1},\hat{x}_\lambda\big]=i\big({-}a_\mu\eta_{\lambda\nu}+ a_\lambda O^{-1}_{\mu\nu}\big),
\qquad
[O_{\mu\nu},\hat{y}_\lambda]=i(a_\mu\eta_{\lambda\nu}- a_\lambda O_{\mu\nu})
\end{gather*}
and
\begin{gather*}
\big[O_{\mu\nu}^{-1},\hat{y}_\lambda\big]=i\big({-}a_\mu O^{-1}_{\lambda\nu}+ a_\lambda
O^{-1}_{\mu\nu}\big)=-iC_{\mu\lambda\alpha}\big(O^{-1}\big)^{\alpha\nu}.
\end{gather*}
The commutation relation $[O_{\mu\nu}^{-1},\hat{x}_\lambda]$ can be also obtained from~\eqref{b18} multiplying by~$O_{\mu\alpha}^{-1}$ and~$O_{\beta\nu}^{-1}$ and using $a^\alpha O_{\mu\alpha}^{-1}=a_\mu$.
Let us mention that elements $O_{\mu\nu}$ satisfy
\begin{gather}
\label{b19}
O_{\mu\nu}=\eta_{\mu\nu}+C_{\ \mu\nu}^{\alpha}p_{\alpha}^{\rm L}.
\end{gather}

One can easily check that $\hat{y}_\mu{\blacktriangleright}1=\hat{x}_\mu$.
Using~\eqref{b17} and~\eqref{b19}, it is easy to obtain that $\hat{y}_\mu{\blacktriangleright}\hat{x}_\nu=
\hat{x}_\nu\hat{x}_\mu$ and
\begin{gather*}
\hat{f}(\hat{y}){\blacktriangleright}\hat{g}(\hat{x})=\hat{g}(\hat{x})\hat{f}^{\rm op}(\hat{x}).
\end{gather*}
Here $\hat{f}^{\rm op}$ stands for the {\it opposite} polynomial ($(\hat{x}_\mu\hat{x}_\nu)^{\rm op}=\hat{x}_\nu\hat{x}_\mu$).
Hence, the action ${\blacktriangleright}$ of $\hat{f}(\hat{y})$ can be understood as a~multiplication from the
right with $\hat{f}^{\rm op}(\hat{x})$.
One can show that $\bigtriangleup(\hat{y}_\mu)=1\otimes\hat{y}_\mu$.
Note that the same construction as for~$\kappa$-Minkowski space~\eqref{b1} could be generalized to arbitrary Lie algebra
def\/ined by structure constants $C_{\mu\nu\lambda}$.

\section{Hopf algebroid}
\label{ha}

\subsection[Hopf algebroid structure of $\hat{\mathcal{H}}$]{Hopf algebroid structure of $\boldsymbol{\hat{\mathcal{H}}}$}
\label{dha}

We def\/ine the source map, target map, coproduct, counit and antipode such that ${\hat{\mathcal{H}}}$ has the
structure of the Hopf algebroid.

In Hopf algebroid, the unit map is replaced by the source and target maps.
In our case ${\hat{\mathcal{H}}}$ is the total algebra and ${\hat{\mathcal A}}$ is the base algebra.
The source map ${\hat{\alpha}}:{\hat{\mathcal A}}\rightarrow{\hat{\mathcal{H}}}$ is
def\/ined~by
\begin{gather*}
%\label{c01}
{\hat{\alpha}}\big(\hat{f}(\hat{x})\big)=\hat{f}(\hat{x}).
\end{gather*}
The target map ${\hat{\beta}}:{\hat{\mathcal A}}\rightarrow{\hat{\mathcal{H}}}$ is
def\/ined~by
\begin{gather*}
%\label{c02}
{\hat{\beta}}\big(\hat{f}(\hat{x})\big)=\hat{f}^{\rm op}(\hat{y}).
\end{gather*}
Let us recall that the source map is the homomorphism while the target map is the antihomomorphism.
Relation~\eqref{b17} shows that
\begin{gather*}
{\hat{\alpha}}\big(\hat{f}(\hat{x})\big){\hat{\beta}}(\hat{g}(\hat{x}))=
{\hat{\beta}}(\hat{g}(\hat{x})){\hat{\alpha}}\big(\hat{f}(\hat{x})\big).
\end{gather*}

In order to def\/ine the coproduct on ${\hat{\mathcal{H}}}$, we consider the subspace
${\hat{\mathcal{B}}}$ of ${\hat{\mathcal{H}}}\otimes{\hat{\mathcal{H}}}$:
\begin{gather*}
{\hat{\mathcal{B}}}=\mathcal{U}\big({\hat{R}_\mu}\big)\big({\hat{\mathcal
A}}\otimes{\mathbb C}\big)\bigtriangleup{\mathcal T},
\end{gather*}
where $\mathcal{U}({\hat{R}_\mu})$ denotes the universal enveloping algebra generated~by
${\hat{R}_\mu}$ (see~\eqref{c03}).
Here, $\bigtriangleup{\mathcal T}$ denotes the subalgebra of
${\hat{\mathcal{H}}}\otimes{\hat{\mathcal{H}}}$ generated by $1\otimes1$ and elements
$\bigtriangleup(p_\mu)$.
For example, we can consider $p_\mu^{\rm L}$ and then $\bigtriangleup{\mathcal T}$ is generated~by
$1\otimes1$ and $p^{\rm L}_\mu\otimes Z^{-1}+1\otimes p^{\rm L}_\mu$.
Since
\begin{gather}
\label{c03}
\big[{\hat{R}_\mu},{\hat{R}_\nu}\big]
=i\big(a_\mu{\hat{R}_\nu}-a_\nu{\hat{R}_\mu}\big)=iC_{\mu\nu\alpha}{\hat{R}_\alpha},
\\
\label{c04}
\big[\hat{x}_\mu\otimes1,{\hat{R}_\nu}\big]=i\big(a_\mu{\hat{R}_\nu}-a_\nu{\hat{R}_\mu}\big),
\\
\label{c05}
\big[O_{\mu\alpha}\otimes\hat{x}^\alpha,{\hat{R}_\nu}\big]=0,
\\
\big[\bigtriangleup p^{\rm L}_\mu,{\hat{R}_\nu}\big]=0
\nonumber
\end{gather}
and
\begin{gather}
\label{c07}
\big[\hat{x}_\mu\otimes1,p_\nu^{\rm L}\otimes Z^{-1}+1\otimes p_\nu^{\rm L}\big]= i\eta_{\mu\nu}Z^{-1}\otimes
Z^{-1}\in\bigtriangleup{\mathcal T},
\end{gather}
${\hat{\mathcal{B}}}$ is a~subalgebra of
${\hat{\mathcal{H}}}\otimes{\hat{\mathcal{H}}}$.
It is obvious that~\eqref{c05} is a~consequence of~\eqref{c03} and~\eqref{c04} but we write it for completeness.
Now, let us consider the subspace ${\hat{\mathfrak{I}}}$ of ${\hat{\mathcal{B}}}$ def\/ined~by
\begin{gather*}
{\hat{\mathfrak{I}}}=\mathcal{U}_{+}\big({\hat{R}_\mu}\big)\big({\hat{\mathcal
A}}\otimes{\mathbb C}\big)\bigtriangleup{\mathcal T},
\end{gather*}
where $\mathcal{U}_{+}({\hat{R}_\mu})$ is the universal enveloping algebra generated~by
${\hat{R}_\mu}$ but without the unit element.
Using~\eqref{c03}--\eqref{c07} one can check that
${\hat{\mathfrak{I}}}={\hat{\mathfrak{K}}}\cap{\hat{\mathcal{B}}}$ and
${\hat{\mathfrak{I}}}$ is the twosided ideal in ${\hat{\mathcal{B}}}$.

\begin{remark*} We could also def\/ine the subalgebra ${\hat{\mathcal{B}}}_3$ in
${\hat{\mathcal{H}}}\otimes{\hat{\mathcal{H}}}\otimes{\hat{\mathcal{H}}}$~by
\begin{gather*}
{\hat{\mathcal{B}}}_3=\mathcal{U}\big[\big({\hat{R}_\mu}\big)_{1,2},\big({\hat{R}_\mu}\big)_{2,3}\big]\big({\hat{\mathcal
A}}\otimes{\mathbb C}\otimes{\mathbb C}\big)(\bigtriangleup\otimes 1)
(\bigtriangleup{\mathcal T}),
\end{gather*}
where $\mathcal{U}[({\hat{R}_\mu})_{1,2},({\hat{R}_\mu})_{2,3}]$
denotes the universal enveloping
algebra generated by $1\otimes1\otimes1$, $({\hat{R}_\mu})_{1,2}={\hat{R}_\mu}\otimes1$
and $({\hat{R}_\mu})_{2,3}=1\otimes{\hat{R}_\mu}$ and we have that
$(\bigtriangleup\otimes 1)(\bigtriangleup{\mathcal
T})=(1\otimes\bigtriangleup)(\bigtriangleup{\mathcal T})$ since~${\mathcal T}$ is a~Hopf algebra.
Similarly, we can def\/ine ${\hat{\mathcal{B}}}_n$ and then ${\hat{\mathcal{B}}}$ would correspond to
${\hat{\mathcal{B}}}_2$.
Also, ${\hat{\mathfrak{K}}}_n$~and
${\hat{\mathfrak{I}}}_n={\hat{\mathfrak{K}}}_n\cap{\hat{\mathcal{B}}}_n$ can be def\/ined.
See~\cite{l95} for the similar discussion.
\end{remark*}

Now, we def\/ine the coproduct
$\bigtriangleup:{\hat{\mathcal{H}}}\rightarrow{\hat{\mathcal{B}}}/{\hat{\mathfrak{I}}}
=\bigtriangleup{\hat{\mathcal{H}}}$
by
\begin{gather}
\label{c1}
\bigtriangleup(\hat{x}_\mu)=\hat{x}_\mu\otimes1+{\hat{\mathfrak{I}}}=
Z^{-1}\otimes\hat{x}_\mu-a_\mu p^{\rm L}_\alpha\otimes\hat{x}_\alpha+{\hat{\mathfrak{I}}}=
O_{\mu\alpha}\otimes\hat{x}_\alpha+{\hat{\mathfrak{I}}},
\\
%\label{c2}
\bigtriangleup\big(p^{\rm L}_\mu\big)=p^{\rm L}_\mu\otimes Z^{-1}+1\otimes p^{\rm L}_\mu+{\hat{\mathfrak{I}}}.
\nonumber
\end{gather}
Notice that ${\hat{\mathcal{B}}}/{\hat{\mathfrak{I}}}$ is the ``restriction'' of Lu's
${\hat{\mathcal{H}}}\otimes{\hat{\mathcal{H}}}/{\hat{\mathfrak{K}}}$, or in other words
an $({\hat{\mathcal A}},{\hat{\mathcal A}})$-submodule of
${\hat{\mathcal{H}}}\otimes{\hat{\mathcal{H}}}/{\hat{\mathfrak{K}}}$ that turns out to
be an algebra, which, in turn, allows us to def\/ine $\bigtriangleup$ as an algebra homomorphism
\begin{gather*}
%\label{c3}
\bigtriangleup\big(\hat{f}\hat{g}\big)=\bigtriangleup\big(\hat{f}\big)\bigtriangleup(\hat{g}).
\end{gather*}
The coproduct of $\hat{y}_\mu$ is given~by
$\bigtriangleup(\hat{y}_\mu)=1\otimes\hat{y}_\mu+{\hat{\mathfrak{I}}}=\hat{y}_\alpha \otimes
O^{-1}_{\mu\alpha}+{\hat{\mathfrak{I}}}$.
One can check that such def\/ined coproduct is coassociative.

The counit ${\hat{\epsilon}}:{\hat{\mathcal{H}}}\rightarrow{\hat{\mathcal A}}$ is
def\/ined~by
\begin{gather*}
%\label{c4}
{\hat{\epsilon}}\big(\hat{h}\big)=\hat{h}{\blacktriangleright}1.
\end{gather*}
This map is not a~homomorphism.
It is easy to check that $m({\hat{\alpha}}{\hat{\epsilon}}\otimes 1)\bigtriangleup=1$
and $m(1\otimes{\hat{\beta}}{\hat{\epsilon}})\bigtriangleup=1$.
In order to check the f\/irst identity, we write elements of ${\hat{\mathcal{H}}}$ in the form
$\hat{f}(\hat{x})g(p)$ and for the second identity in the form $\hat{f}(\hat{y})g(p)$.

The antipode ${S}:{\hat{\mathcal{H}}}\rightarrow{\hat{\mathcal{H}}}$ is def\/ined~by
\begin{gather*}
%\label{c7}
{S}(\hat{y}_\mu)=\hat{x}_\mu
\qquad
\text{and}
\qquad
%\label{c10}
{S}\big(p^{\rm L}_\mu\big)=-p^{\rm L}_\mu Z.
\end{gather*}
The antipode ${S}(\hat{x}_\mu)$ can be calculated from~\eqref{b16}.
One obtains that
\begin{gather}
\label{c11}
{S}(\hat{x}_\mu)=\hat{y}_\mu+ia_\mu(1-n).
\end{gather}
It follows that ${S}^2(\hat{y}_\mu)=\hat{y}_\mu+ia_\mu(1-n)$ (and
${S}^2(\hat{x}_\mu)=\hat{x}_\mu+ia_\mu(1-n)$) and ${S}^2(p_\mu)=p_\mu$.
Previous two formulas can be written also as ${S}^2(\hat{h})=Z^{1-n}\hat{h}Z^{n-1}$.
It is enough to check it for the elements $\hat{x}_\mu$ and $p_\mu$ since $S^2$ is a~homomorphism.
The expression of $S^2(\hat{x}_\mu)$ can be written in terms of structure constants:
\begin{gather}
\label{c12}
S^2(\hat{x}_\mu)=\hat{x}_\mu+iC_{\alpha\mu}^{\ \ \alpha}.
\end{gather}

A nice way to check the consistency of the antipode is to start with~\eqref{b16} and apply the antipode~$S$ (note that
${S}(O_{\mu\alpha})=O_{\mu\alpha}^{-1}$):
\begin{gather*}
{S}(\hat{x}_\mu)=O_{\mu\alpha}^{-1}\hat{x}^\alpha=O_{\mu\alpha}^{-1} \hat{y}_\beta O^{\alpha\beta}.
\end{gather*}
It produces
\begin{gather*}
{S}^2(\hat{x}_\mu)=\big(O^{-1}\big)^{\alpha\beta}\hat{x}_\beta O_{\mu\alpha}.
\end{gather*}
It remains to apply expressions for $(O^{-1})^{\alpha\beta}$ and $O_{\mu\alpha}$ (see~\eqref{b13} and~\eqref{b11}), use
the abbreviation $A^{\rm L}=-a^\alpha p_\alpha^{\rm L}=-(ap^{\rm L})$ and recall the identity $Z=(1-A^{\rm L})^{-1}$ (see~\cite{km11}).

Let $\mathcal{P}\subset{\hat{\mathcal{H}}}$ be the enveloping algebra of the Poincar\'{e} algebra
${\mathfrak p}$.
It is possible to def\/ine the Hopf algebra structure on the subalgebra $\mathcal{P}$~\cite{km11}.
It is interesting to note that the coproduct and the antipode map def\/ined above on ${\hat{\mathcal{H}}}$ and
restricted to $\mathcal{P}$ coincides with the coproduct and the antipode map on the Hopf algebra
$\mathcal{P}$~\cite{jms13b}.
For more details see Section~\ref{Section5}.

It is easy to check that
\begin{gather}
\label{nova}
{S}{\hat{\beta}}={\hat{\alpha}},
\qquad
m(1\otimes{S})\bigtriangleup={\hat{\alpha}}{\hat{\epsilon}},
\qquad
m({S}\otimes 1)\bigtriangleup={\hat{\beta}}{\hat{\epsilon}}{S}.
\end{gather}
The f\/irst identity is obvious, the second one can be easily checked for the base elements and the third identity can be
easily checked using the dual basis.

In~\cite{l95}, Lu analyzes the right ideal ${\hat{\mathfrak{K}}}$ generated~by
${\hat{Q}_\mu}=\hat{y}_\mu\otimes1-1\otimes\hat{x}_\mu$ (right ideal ${\hat{\mathfrak{K}}}$ is
denoted by $I_2$ in~\cite{l95}).
These elements are equal to ${\hat{R}_\alpha}((O^{-1})^{\mu\alpha}\otimes1)$.
It is important to mention that the identity
$m(1\otimes{S})\bigtriangleup={\hat{\alpha}}{\hat{\epsilon}}$ is not
satisf\/ied in~\cite{l95}, because $m(1\otimes {S}){\hat{\mathfrak{K}}}\neq0$ and this is why the
section~$\gamma$ is needed.
In our approach, since we have
$\bigtriangleup:{\hat{\mathcal{H}}}\rightarrow{\hat{\mathcal{B}}}/{\hat{\mathfrak{I}}}
=\bigtriangleup{\hat{\mathcal{H}}}$
and
\begin{gather*}
m(1\otimes{S}){\hat{\mathfrak{I}}}=0,
\end{gather*}
it is easy to see that~\eqref{nova} holds $\forall\, h\in{\hat{\mathcal{H}}}$.

Let us point out that $[{\hat{R}_\mu},{\hat{Q}_\nu}]=0$ and
$[{\hat{Q}_\mu},{\hat{Q}_\nu}]=i(-a_\mu{\hat{Q}_\nu}+a_\nu{\hat{Q}_\mu})$.
Also, it is easy to check that $[{\hat{Q}_\mu},\bigtriangleup p^{\rm L}_\nu]=0$ and
$[{\hat{Q}_\mu},\hat{x}_\nu\otimes1]=0$.

\subsection[Hopf algebroid structure of $\mathcal{H}$]{Hopf algebroid structure of $\boldsymbol{\mathcal{H}}$}\label{uha}

Now, let us consider the case when the deformation vector $a_{\mu}$ is equal to~0.
Then~\eqref{b1} transforms~to
\begin{gather*}
%\label{c33}
[\hat{x}_\mu,\hat{x}_\nu]=0,
\end{gather*}
the algebra ${\hat{\mathcal{H}}}$ becomes the Weyl algebra which we denote by ${\mathcal{H}}$ and
write $x_\mu$ instead of $\hat{x}_\mu$.
We have already mentioned that it is not possible to construct the Hopf algebra structure on ${\mathcal{H}}$.
Let us repeat the Hopf algebroid structure on ${\mathcal{H}}$ and set the terminology.

Now, $\varphi_{\mu\nu}=O_{\mu\nu}=\eta_{\mu\nu}$, $Z=1$ and $\hat{y}_\mu=x_\mu$.
Let $\mathcal A$\ (the base algebra) be the subalgebra of~${\mathcal{H}}$ generated by~1 and~$x_\mu$.
We def\/ine the action $\rhd$ of ${\mathcal{H}}$ on $\mathcal A$ in the same way as we did it in
Section~\ref{ps}: $f(x)\rhd g(x)=f(x)g(x)$, $p_\mu\rhd1=0$ and $p_\mu\rhd g(x)=[p_\mu,g(x)]\rhd1=p_\mu g(x)\rhd1$.
Then~$\mathcal A$ can be considered as an ${\mathcal{H}}$-module.
It is clear that the action ${\blacktriangleright}$ transforms to the action~$\rhd$ when the vector~$a$ is equal to~0.

The source and the target map are now equal ${\alpha}_0={\beta}_0$ and $\alpha_0; \beta_0:
\mathcal{A}\rightarrow\mathcal{H} $ reduces to the natural inclusion.

The counit ${\epsilon_0}:{\mathcal{H}}\rightarrow{\mathcal A}$ is def\/ined~by
\begin{gather*}
%\label{c36}
{\epsilon_0}(h)=h\rhd1.
\end{gather*}
In order to def\/ine the coproduct, let us def\/ine relations $(R_0)_\mu$~by
\begin{gather*}
(R_0)_\mu=x_\mu\otimes1-1\otimes x_\mu.
\end{gather*}
Let $\mathcal{U}[(R_0)_\mu]$ be the universal enveloping algebra generated by $1\otimes1$ and $(R_0)_\mu$,
$\mathcal{U}_+[(R_0)_\mu]$ be the universal enveloping algebra generated by $(R_0)_\mu$ but without the unit element,
and $\bigtriangleup_0{\mathcal T}$ be the algebra generated by $1\otimes1$ and
$p_\mu\otimes1+1\otimes p_\mu$.
Note that ${\mathcal T}$ is isomorphic to $\bigtriangleup_0{\mathcal T}$.
Now, we def\/ine ${\mathcal{B}_0}$, the subalgebra of ${\mathcal{H}}\otimes{\mathcal{H}}$ of the form
\begin{gather*}
{\mathcal{B}_0}=\mathcal{U}[(R_0)_\mu]({\mathcal A}\otimes{\mathbb
C})\bigtriangleup_0{\mathcal T}
\end{gather*}
and twosided ideal ${\mathfrak{I}_0}$ of ${\mathcal{B}_0}$~by
\begin{gather*}
{\mathfrak{I}_0}=\mathcal{U}_+[(R_0)_\mu]({\mathcal A}\otimes{\mathbb
C})\bigtriangleup_0{\mathcal T}.
\end{gather*}
The coproduct
$\bigtriangleup_0:{\mathcal{H}}\rightarrow{\mathcal{B}_0}/{\mathfrak{I}_0}
=\bigtriangleup_0{\mathcal{H}}$
is a~homomorphism def\/ined~by
\begin{gather*}
\bigtriangleup_0(x_\mu)=x_\mu\otimes1+{\mathfrak{I}_0},
\qquad
\bigtriangleup_0(p_\mu)=p_\mu\otimes1+1\otimes p_\mu+{\mathfrak{I}_0}.
\end{gather*}
One checks that the coproduct $\bigtriangleup_0$ and the counit ${\epsilon_0}$ satisfy
$m({\alpha}_0{\epsilon_0}\otimes 1)\bigtriangleup_0=1$ and
$m(1\otimes{\beta}_0{\epsilon_0})\bigtriangleup_0=1$.

The antipode $S_0:{\mathcal{H}}\rightarrow{\mathcal{H}}$ transforms to
\begin{gather}
\label{d7}
S_0(x_\mu)=x_\mu,
\qquad
S_0(p_\mu)=-p_\mu.
\end{gather}
It is easy to check that
\begin{gather}
\label{nova1}
m(1\otimes S_0)\bigtriangleup_0={\alpha}_0{\epsilon_0},
\qquad
m(S_0\otimes 1)\bigtriangleup_0={\beta}_0{\epsilon_0} S_0.
\end{gather}
Similarly as in the deformed case, the expression $m(1\otimes S_0)\bigtriangleup_0$ is not well def\/ined
in~\cite{l95}, because $m(1\otimes S_0)\mathcal{K}_0\neq0$ and this is why the section~$\gamma$ is needed.
In our approach, since
\begin{gather*}
m(1\otimes S_0){\mathfrak{I}_0}=0
\end{gather*}
holds, one can check~\eqref{nova1} $\forall\, h\in{\mathcal{H}}$.

\section{Twisting Hopf algebroid structure}
\label{tw}

\subsection{Realizations}
%\label{re}

The phase space satisfying~\eqref{b1} and~\eqref{b4} can be analyzed by realizations (see~\cite{km11,hermitian, mssg08}).
In Section~\ref{uha}, we have analyzed the Weyl algebra ${\mathcal{H}}$ generated by $p_\mu$ and
commutative coordinates $x_\mu$ satisfying
\begin{gather*}
%\label{d1}
[p_\mu,x_\nu]=-i\eta_{\mu\nu}1.
\end{gather*}
Then, the noncommutative coordinates $\hat{x}_\mu$ are expressed in the form
\begin{gather}
\label{d4}
\hat{x}_\mu=x^\alpha \varphi_{\alpha\mu}(p)
\end{gather}
such that~\eqref{b1} and~\eqref{b4} are satisf\/ied.
It is important to observe that the space ${\mathcal{H}}$ is isomorphic to ${\hat{\mathcal{H}}}$ as an algebra.
Hence, we set ${\hat{\mathcal{H}}}= {\mathcal{H}}$ and treat sets $\{x_\mu,p_\nu\}$ and
$\{\hat{x}_\mu,p_\nu\}$ as dif\/ferent bases of the same algebra.
However, we will use both symbols, ${\hat{\mathcal{H}}}$ and ${\mathcal{H}}$ in order to emphasize the basis.
The action $\rhd$, def\/ined in Section~\ref{uha} corresponds to ${\mathcal{H}}$.
However, ${\mathcal{H}}$ and ${\hat{\mathcal{H}}}$, considered as Hopf algebroids are dif\/ferent.

The restriction of the counit ${\epsilon_0}|_{{\hat{\mathcal A}}}$, introduced in
Section~\ref{uha}, def\/ines the bijection of vector spaces ${\hat{\mathcal A}}$ and ${\mathcal A}$.
By the abuse of notation, we denote it by ${\epsilon_0}$ or $\rhd$.
Let us mention that the inverse map is simply ${\hat{\epsilon}}|_{{\mathcal A}}$.
Then, the star product $\star$ on $\mathcal A$\ is def\/ined
by $(f\star g)(x)=\hat{f}(\hat{x})\hat{g}(\hat{x})\rhd1= \hat{f}(\hat{x})\rhd g(x)$ where $f=\hat{f}\rhd1$ and $g=\hat{g}\rhd1$.
The algebra $\mathcal A$\ equipped with the star product instead of pointwise multiplication will be denoted
by ${{\mathcal A}_\star}$ and
the map ${\epsilon_0}:{\hat{\mathcal A}}\rightarrow{{\mathcal A}_\star}$ is an isomorphism of algebras.

It is possible to construct the dual realization $\tilde{\varphi}_{\mu\nu}$ and the dual star product
$\star_{\tilde{\varphi}}$ such that
\begin{gather*}
(f\star_\varphi g)(x)=(g\star_{\tilde{\varphi}}f)(x)
\end{gather*}
is satisf\/ied (see~\cite[Section 5]{km11}).
Now, elements $\hat{y}_\mu$ are given~by
\begin{gather*}
\hat{y}_\mu=x^\alpha\tilde{\varphi}_{\alpha\mu}(p).
\end{gather*}
It is easy to check the following properties:
\begin{gather*}
\hat{x}_\mu\rhd f(x)=x_\mu\star_\varphi f(x)=f(x)\star_{\tilde{\varphi}}x_\mu
\end{gather*}
and
\begin{gather*}
\hat{y}_\mu\rhd f(x)=x_\mu\star_{\tilde{\varphi}} f(x)=f(x)\star_\varphi x_\mu.
\end{gather*}

\subsubsection{Similarity transformations}

The relation between realizations is given by the similarity transformations~\cite{jms13a}.
Let us consider two realizations.
The f\/irst one is denoted by $x_\mu$ and $p_\mu$ and given by the set of functions $\{\varphi_{\mu\nu}\}$ (and~\eqref{b4}
or~\eqref{d4}).
The second realization is denoted by $X_\mu$, $P_\mu$ and $\Phi_{\mu\nu}$ ($\hat{x}_\mu=X^\alpha \Phi_{\alpha\mu}(P)$).
The similarity transformation ${{\mathcal E}}$ is given
by ${{\mathcal E}}=\exp\{x^\alpha\Sigma_\alpha(p)\}$ such that $\lim\limits_{a\rightarrow0}\Sigma_\alpha=0$.
Now, the relation between realizations is given~by
\begin{gather*}
P_\mu = {{\mathcal E}} p_\mu{{\mathcal E}}^{-1},
\qquad
X_\mu = {{\mathcal E}} x_\mu{{\mathcal E}}^{-1}.
\end{gather*}
It is easy to see that $P_\mu=P_\mu(p)$.
Since
$[P_\mu,\hat{x}_\nu]=-i\Phi_{\mu\nu}(P)$,
\begin{gather*}
\frac{\partial P_\mu}{\partial p_\alpha}\varphi_{\alpha\nu}=\Phi_{\mu\nu}(P(p))
\qquad \text{and} \qquad
\varphi_{\alpha\nu}=\left[\frac{\partial P}{\partial p}\right]_{\alpha\mu}^{-1} \Phi_{\mu\nu}(P(p)).
\end{gather*}
It follows that the set of functions $\varphi_{\mu\nu}$ can be obtained from the set of functions $\Phi_{\mu\nu}$ and
the expressions of~$P$ in terms of~$p$.
Since $O_{\mu\nu}=O_{\mu\nu}(P(p))$, it is easy to express $O_{\mu\nu}$ in the realization determined by $x_\mu$ and
$p_\mu$.

\subsubsection{Examples}

Let us consider three examples of realizations.
The {\it noncovariant~$\lambda$-family} of realizations is given~by
\begin{gather}
\label{d16}
\hat{x}_{0}=x_{0}^{(\lambda)}-a_{0}\left(1-\lambda \right) x_{k}^{(\lambda)}p_{k}^{(\lambda)},
\qquad
\hat{x}_{k}=x_{k}^{(\lambda)}Z^{-\lambda},
\end{gather}
and
\begin{gather}
\label{d17}
\hat{y}_0=\hat{x}_0Z-ia_0+a_0\big(\hat{x}p^{\rm L}\big)Z,
\qquad
\hat{y}_j=\hat{x}_jZ,
\end{gather}
where $Z=e^{A^{(\lambda)}}$ and $\lambda\in{\mathbb R}$.
For this family we assume that $a=(a_0,0,\ldots,0)$.
Here, $(\lambda)$ denotes the label.
Generic realizations are denoted without the label.
It is easy to obtain $p_0^{\rm L}=\frac1{a_0}(1-Z^{-1})$ and $p_k^{\rm L}=p_k^{(\lambda)} Z^{\lambda-1}$.
Now, one calculates $O_{\mu\nu}$ (see~\eqref{b11}) in terms of $p_\mu^{(\lambda)}$: $O_{k\nu}=Z^{-1}\eta_{k\nu}$,
$O_{00}=-1$ and
\begin{gather*}
O_{0k}=Z^{-1}\eta_{0k}-a_0 p^{\rm L}_k=\big(\eta_{0k}- a_0 p_k^{(\lambda)}Z^\lambda\big)Z^{-1}.
\end{gather*}

The {\it left covariant} realization is def\/ined~by
\begin{gather*}
\hat{x}_\mu=x^{\rm L}_\mu\big(1-A^{\rm L}\big),
\end{gather*}
where $Z=(1-A^{\rm L})^{-1}$.
The element $p^{\rm L}$ that we have mentioned in Section~\ref{ps} corresponds to the {\it left covariant} realization.
It is easy to obtain that
\begin{gather*}
\hat{y}_\mu=x_\mu^{\rm L}+a_\mu\big(x^{\rm L}p^{\rm L}\big)
\end{gather*}
(see~\eqref{b12} for the def\/inition of $\hat{y}_\mu$).

The {\it right covariant} realization is def\/ined~by
\begin{gather*}
\hat{x}_\mu=x_\mu^{\rm R}-a_\mu\big(x^{\rm R}p^{\rm R}\big),
\end{gather*}
where $Z=1+A^{\rm R}$.
The relation between $p_\mu^{\rm L}$ and $p_\mu^{\rm R}$ is given by $p_\mu^{\rm R}=p_\mu^{\rm L}Z$.
Now,
\begin{gather*}
\hat{y}_\mu=x_\mu^{\rm R}\big(1+A^{\rm R}\big).
\end{gather*}
Also, it easy to calculate $O_{\mu\nu}$ in terms of $p_\mu^{\rm R}$:
\begin{gather*}
O_{\mu\nu}=Z^{-1}\eta_{\mu\nu}-a_\mu p^{\rm L}_\nu=\big(\eta_{\mu\nu}- a_\mu p_\nu^{\rm R}\big)Z^{-1}.
\end{gather*}

One should notice the duality between the {\it left covariant} and the {\it right covariant} realizations.

\subsection{Twist and Hopf algebroid}

For each realization, there is the corresponding twist and vice versa~\cite{jms13a}.
The relation between the star product and twist is given~by
\begin{gather*}
f\star g=m\big(\mathcal{F}^{-1}\rhd(f\otimes g)\big)
\end{gather*}
for $f,g\in{\mathcal A}$.
It follows that
$\mathcal{F}^{-1}\in{\mathcal{H}}\otimes{\mathcal{H}}/{\mathfrak{K}}_0$.
Now, we will use twists to reconstruct the Hopf algebroid structure described in Section~\ref{dha}, from the Hopf
algebroid structure analyzed in Section~\ref{uha}.
That is we will show that by twisting the Hopf algebroid structure of $\mathcal{H}$ one can obtain the Hopf algebroid
structure of $\hat{\mathcal{H}}$.
Hence, we will consider twists $\mathcal{F}$ such that
$\mathcal{F}:\bigtriangleup_0{\mathcal{H}}\rightarrow\bigtriangleup{\mathcal{H}}$.
Here ${\mathfrak{I}}\cong{\hat{\mathfrak{I}}}$ and
$\bigtriangleup{\mathcal{H}}\cong\bigtriangleup{\hat{\mathcal{H}}}$.
More precisely, ${\mathfrak{I}}$ is the twosided ideal generated by elements~$R_\mu$ which are def\/ined~by
\begin{gather*}
%\label{e2}
R_\mu=\mathcal{F}(R_0)_\mu\mathcal{F}^{-1}.
\end{gather*}
Let us mention that the relation between ${\hat{R}_\mu}$ and $R_\mu$ is given~by
\begin{gather*}
{\hat{R}_\mu}=R^\alpha\bigtriangleup(\varphi_{\alpha\mu}).
\end{gather*}
Also, it is easy to rebuild the realization from the twist.
For the given twist $\mathcal{F}$, the corresponding realization is obtained~by
\begin{gather*}
%\label{e7}
\hat{x}_\mu=m\big(\mathcal{F}^{-1}(\rhd\otimes 1)(x_\mu\otimes1)\big).
\end{gather*}
Similarly,
\begin{gather*}
%\label{e8}
\hat{y}_\mu=m\big(\tilde{\mathcal{F}}^{-1}(\rhd\otimes 1) (x_\mu\otimes1)\big),
\end{gather*}
where $\tilde{\mathcal{F}}^{-1}$ is given by $\tilde{\mathcal{F}}^{-1}= \tau_0\mathcal{F}^{-1}\tau_0$ ($\tau_{0}$ stands
for the f\/lip operator with the property $\tau_0(h_{1}\otimes h_{2})=h_{2}\otimes h_{1}$, $\forall\, h_{1},h_{2}\in\mathcal{H}$).

The {\it noncovariant~$\lambda$-family} of realizations have twists of the form
\begin{gather}
\label{e1}
\mathcal{F}^{(\lambda)}=\exp \big(i\big(\lambda x_k^{(\lambda)}p_k^{(\lambda)}\otimes A^{(\lambda)}-
(1-\lambda)A^{(\lambda)}\otimes x_k^{(\lambda)}p_k^{(\lambda)}\big)\big).
\end{gather}
These twists belong to the family of Abelian twists (see~\cite{gghmm0809}).
The {\it left covariant} and the {\it right covariant} realizations, respectively, have twists of the form
\begin{gather*}
\mathcal{F}^{\rm L}=\exp\big(i\big(x^{\rm L}p^{\rm L}\big)\otimes\ln Z\big)
\qquad
\text{and}
\qquad
\mathcal{F}^{\rm R}=\exp\big({-}\ln Z\otimes i\big(x^{\rm R}p^{\rm R}\big)\big).
\end{gather*}
These two twists belong to the family of Jordanian twists (see~\cite{bp09}).

Let us reconstruct the source and the target maps from the twist.
First, we def\/ine ${\alpha}$ and ${\beta}$,
${\alpha}:{{\mathcal A}_\star}\rightarrow\mathcal{U}(\hat{x}_\mu)\subset{\mathcal{H}}$,
${\beta}:{{\mathcal A}_\star}\rightarrow\mathcal{U}(\hat{y}_\mu)\subset{\mathcal{H}}$~by
\begin{gather*}
{\alpha}(f(x))=m\big(\mathcal{F}^{-1}(\rhd\otimes 1) ({\alpha}_0(f(x))\otimes1)\big),
\qquad
{\alpha}_0(f(x))=f(x),
\end{gather*}
and
\begin{gather*}
{\beta}(f(x))=m\big(\tilde{\mathcal{F}}^{-1}(\rhd\otimes 1) ({\beta}_0(f(x))\otimes1)\big),
\qquad
{\beta}_0(f(x))=f(x).
\end{gather*}
Now, the source and the target maps are given~by
\begin{gather*}
{\hat{\alpha}}={\alpha}{\epsilon_0}|_{{\hat{\mathcal A}}}
\qquad
\text{and}
\qquad
{\hat{\beta}}={\beta}{\epsilon_0}|_{{\hat{\mathcal A}}}.
\end{gather*}
The counit ${\hat{\epsilon}}:{\mathcal{H}}\rightarrow{\hat{\mathcal A}}$ is given~by
\begin{gather*}
{\hat{\epsilon}}(h)=m\big(\mathcal{F}^{-1}(\rhd\otimes 1)({\epsilon_0}(h)\otimes1)\big).
\end{gather*}
The coproduct can be calculated by the formula:
\begin{gather*}
\bigtriangleup(h)=\mathcal{F}(\bigtriangleup_0(h))\mathcal{F}^{-1}.
\end{gather*}

For the {\it noncovariant~$\lambda$-family} of realizations
\begin{gather}
\label{e13}
\bigtriangleup\big(x_j^{(\lambda)}\big)=x_j^{(\lambda)}\otimes Z^\lambda=Z^{\lambda-1} \otimes x_j^{(\lambda)},
\\
%\label{e14}
\bigtriangleup\big(x_0^{(\lambda)}\big)=x_0^{(\lambda)}\otimes1+a_0(1-\lambda) \otimes
x_k^{(\lambda)}p_k^{(\lambda)}= 1\otimes x_0^{(\lambda)}-a_0\lambda x_k^{(\lambda)}p_k^{(\lambda)}\otimes1,
%\nonumber
\\
%\label{e15}
\bigtriangleup\big(p_j^{(\lambda)}\big)=p_j^{(\lambda)}\otimes Z^{-\lambda}+ Z^{1-\lambda}\otimes p_j^{(\lambda)},
%\nonumber
\end{gather}
and
\begin{gather}
\label{e16}
\bigtriangleup\big(p_0^{(\lambda)}\big)=p_0^{(\lambda)}\otimes1+1\otimes p_0^{(\lambda)}.
\end{gather}
It is a~nice exercise to express $\hat{x}_\mu$ in terms of $x_\alpha^{(\lambda)}$ and $p_\alpha^{(\lambda)}$
(see~\eqref{d16}), use~\eqref{e13}--\eqref{e16} and obtain~\eqref{c1}.

Similarly,
\begin{gather*}
\bigtriangleup\big(x_\mu^{\rm L}\big)=x_\mu^{\rm L}\otimes Z=1\otimes\big(x_\mu^{\rm L}+ia_\mu Z\big)
\qquad
\text{and}
\qquad
\bigtriangleup\big(p_\mu^{\rm L}\big)=p_\mu^{\rm L}\otimes Z^{-1}+1\otimes p_\mu^{\rm L}
\end{gather*}
for the {\it left covariant} realization and
\begin{gather*}
\bigtriangleup\big(x_\mu^{\rm R}\big)=\big(x_\mu^{\rm R}-ia_\mu Z\big)\otimes1=Z^{-1}\otimes x_\mu^{\rm R}
\qquad
\text{and}
\qquad
\bigtriangleup\big(p_\mu^{\rm R}\big)=p_\mu^{\rm R}\otimes1+Z\otimes p_\mu^{\rm R}
\end{gather*}
for the {\it right covariant} realization.

It remains to consider the antipode.
Let
\begin{gather*}
%\label{e3}
\chi^{-1}=m(S_0\otimes 1)\mathcal{F}^{-1},
\end{gather*}
then
\begin{gather}
\label{e4}
S(h)=\chi(S_0(h))\chi^{-1}
\end{gather}
where $S_0$ denotes the undeformed antipode map def\/ined by~\eqref{d7} ($S_0(x_\mu)=x_\mu$ and $S_0(p_\mu)=-p_\mu$).
For the similar approach regarding Hopf algebras, see~\cite{admw05, admw051}.

For the {\it noncovariant~$\lambda$-family} of realizations,~$\chi$ has the form
\begin{gather*}
\chi^{(\lambda)}=\exp \big(i(1-2\lambda)A^{(\lambda)}x_k^{(\lambda)}
p_k^{(\lambda)}+\lambda(1-n)A^{(\lambda)}\big).
\end{gather*}
Then
\begin{gather}
\label{e17}
S\big(p_j^{(\lambda)}\big)=-p_j^{(\lambda)}Z^{2\lambda-1},
\\
%\label{e18}
S\big(p_0^{(\lambda)}\big)=-p_0^{(\lambda)},
 %\nonumber
\\
%\label{e19}
S\big(x_j^{(\lambda)}\big)=x_j^{(\lambda)}Z^{1-2\lambda},
 %\nonumber
\\
\label{e20}
S\big(x_0^{(\lambda)}\big)=x_0^{(\lambda)}-(1-2\lambda) a_0x_k^{(\lambda)}p_k^{(\lambda)}+\lambda ia_0(1-n).
\end{gather}
Again, it is an exercise to express $\hat{x}_\mu$ in terms of $x_\alpha^{(\lambda)}$ and $p_\alpha^{(\lambda)}$
(see~\eqref{d16}), use~\eqref{e17}--\eqref{e20} and obtain~\eqref{c11}.
The antipode is given~by
\begin{gather}
\label{e21}
S(\hat{x}_j)=\hat{x}_jZ
\end{gather}
and
\begin{gather}
\label{e22}
S(\hat{x}_0)=\hat{x}_0+a_0x_k^{(\lambda)}p_k^{(\lambda)}+ia_0(1-n).
\end{gather}
Let us recall that for the {\it noncovariant~$\lambda$-family} of realizations we set $a_\mu=(a_{0},0,...,0)$.
Now, one can compare~\eqref{e21} and~\eqref{e22} with~\eqref{c11}.
The formula for the antipode of $\hat{x}_\mu$ can be also obtained from the formula $S(\hat{y}_\mu)=\hat{x}_\mu$,
formulas for the realization of $\hat{x}_\mu$ and $\hat{y}_\mu$,~\eqref{d16} and~\eqref{d17} and formulas for
$S(p_\mu)$.
For all examples, it is easy to check that $S(\hat{y}_\mu)=\chi(S_0(\hat{y}_\mu))\chi^{-1}=\hat{x}_\mu$.

For the {\it left covariant} realization
\begin{gather*}
\big(\chi^{\rm L}\big)^{-1}=\exp\big(i\big(p^{\rm L}x^{\rm L}\big)A^{\rm L}\big).
\end{gather*}
For the {\it right covariant} realization
\begin{gather*}
\big(\chi^{\rm R}\big)^{-1}=\exp\big({-}iA^{\rm R}\big(x^{\rm R}p^{\rm R}\big)\big).
\end{gather*}

There is a~natural question if the antipode map on the Hopf algebroid ${\hat{\mathcal{H}}}$  def\/ined
by~\eqref{e4} and the antipode map def\/ined on the Hopf algebra ${\mathcal U}(\mathfrak{igl}(n))$ coincide
(see~\cite{klry09} for the formulas of the antipode).
They coincide for $h\in{\hat{\mathcal{H}}}$ for which $\alpha\epsilon (h)=\beta\epsilon S_0(h)$.
For elements~$h$ for which $\alpha\epsilon (h)\neq\beta\epsilon S_0(h)$, the antipode maps do not coincide.
For example, $S_0(x_{j}p_{j})=-x_{j}p_{j}+i$ in the Hopf algebroid, while $S_0(x_{j}p_{j})=-x_{j}p_{j}$ in the Hopf
algebra (here no summation is assumed).
See also~\cite{jms13b}.

Using~\eqref{e4}, it is easy to obtain the expression for $S^{-1}$:
\begin{gather*}
S^{-1}(h)=S_0(\chi)S_0(h)S_0\big(\chi^{-1}\big).
\end{gather*}
One can show that $S_0(\chi)=Z^{n-1}\chi$.
Then $S^{-1}(h)=Z^{n-1}S(h)Z^{1-n}$ and $S^2(h)=Z^{1-n}hZ^{n-1}$.
For example, $S^2(p_\mu)=p_\mu$, $S^2(\hat{x}_\mu)=\hat{x}_\mu+ia_\mu(1-n)$ and
$S^2(\hat{y}_\mu)=\hat{y}_\mu+ia_\mu(1-n)$.
This coincides with results in Section~\ref{ha} (see~\eqref{c11} and~\eqref{c12}).

\section[$\kappa$-Poincar\'e Hopf algebra from~$\kappa$-deformed phase space and twists]{$\boldsymbol{\kappa}$-Poincar\'e
Hopf algebra from~$\boldsymbol{\kappa}$-deformed phase space\\ and twists}
\label{Section5}

Let us consider the~$\kappa$-Poincar\'e Hopf algebra in \textit{natural} realization~\cite{mks07, ms11, mssg08} (or
classical basis~\cite{natbor, Kowalski-Glikman-1}).
We start with the undeformed Poincar\'e algebra generated by Lorentz genera\-tors~$M_{\mu\nu}$ and translation generators
(momentum) $P_{\mu}$
\begin{gather*}
%\label{lorentz}
[M_{\mu\nu},M_{\lambda\rho}]=\eta_{\nu\lambda}M_{\mu\rho}-\eta_{\mu\lambda}M_{\nu\rho}-\eta_{\nu\rho}M_{\mu\lambda}+\eta_{\mu\rho}M_{\nu\lambda},
\\
[P_{\mu},P_{\nu}]=0,
\qquad
[M_{\mu\nu},P_{\lambda}]=\eta_{\nu\lambda}P_{\mu}-\eta_{\mu\lambda}P_{\nu}.
\end{gather*}

The corresponding~$\kappa$-deformed Poincar\'{e}--Hopf algebra can be written in a~unif\/ied covariant way~\cite{Govindarajan-2, remarks,km11, mks07, mssg08}.
The coproduct $\bigtriangleup$ is given~by
\begin{gather}
\bigtriangleup P_{\mu} =P_{\mu}\otimes Z^{-1}+1\otimes P_{\mu}-a_{\mu}p^{\rm L}_{\alpha}Z\otimes P^{\alpha},
\nonumber
\\
\bigtriangleup M_{\mu\nu} =M_{\mu\nu}\otimes 1+1\otimes M_{\mu\nu}-a_{\mu}\big(p^{\rm L}\big)^\alpha Z\otimes
M_{\alpha\nu}+a_{\nu}\big(p^{\rm L}\big)^{\alpha}Z\otimes M_{\alpha\mu},
\label{coproduct}
\end{gather}
as well as the antipode~$S$ and counit~$\epsilon$
\begin{gather}
 S(P_{\mu})=\big({-}P_{\mu}-a_{\mu}p^{\rm L}_{\alpha}P^{\alpha}\big)Z,
\nonumber
\\
 S(M_{\mu\nu})=-M_{\mu\nu} -a_{\mu}\big(p^{\rm L}\big)^\alpha M_{\alpha\nu}+a_{\nu}\big(p^{\rm L}\big)^\alpha M_{\alpha\mu},
\nonumber
\\
 \epsilon(P_{\mu})=\epsilon(M_{\mu\nu})=0,
\label{antipod}
\end{gather}
where the momentum $P_{\mu}$ is related to $p^{\rm L}_{\mu}$ via $P_{\mu}=p^{\rm L}_{\mu}-\frac{a_{\mu}}{2}(p^{\rm L})^2Z$.
The above Hopf algebra structure unif\/ies all three types of deformations $a_{\mu}$, i.e.~time-like ($a^2<0$), space-like
($a^2>0$) and light-like ($a^2=0$).

Using the action $\blacktriangleright$ and coproduct $\bigtriangleup$ we can get the whole algebra
$\left\{\hat{x}_{\mu}, M_{\mu\nu}, P_{\mu}\right\}$ (for details see~\cite{jms2013, jms13b})
\begin{gather}
%\label{Mx}
[M_{\mu\nu},\hat{x}_{\lambda}]=\eta_{\nu\lambda}\hat{x}_{\mu}-\eta_{\mu\lambda}\hat{x}_{\nu}-ia_{\mu}M_{\nu\lambda}+ia_{\nu}M_{\mu\lambda},
\nonumber
\\
\label{Phatx}
[P_{\mu},\hat{x}_{\nu}]=-i\big(\eta_{\mu\nu}Z^{-1}-a_{\mu}P_{\nu}\big),
\end{gather}
where $Z^{-1}=(a P)+\sqrt{1+a^2P^2}$ and from~\eqref{Phatx} it follows that the NC coordinates $\hat{x}_{\mu}$ can be
written in terms of canonical $X_{\alpha}$ and $P_{\alpha}$ ($[X_{\alpha},X_{\beta}]=0$,
$[P_{\mu},X_{\nu}]=-i\eta_{\mu\nu} 1$) via $\hat{x}_{\mu}=X_{\mu}Z^{-1}-(a X)P_{\mu}$ and satisf\/ies~\eqref{b1}.

Now we will discuss the realization of~$\kappa$-Poincar\'e--Hopf algebra via phase space $\hat{\mathcal{H}}$ and discuss
the issue of the twist in the Hopf algebroid approach.
Realization of $M_{\mu\nu}$ in terms of canoni\-cal~$X_{\alpha}$ and~$P_{\alpha}$ is given~by
$M_{\mu\nu}=i(X_{\mu}P_{\nu}-X_{\nu}P_{\mu})$ which for~$\kappa$-deformed phase space variab\-les~$\hat{x}_{\mu}$,~$P_{\mu}$ reads
\begin{gather*}
M_{\mu\nu}=i(\hat{x}_{\mu}P_{\nu}-\hat{x}_{\nu}P_{\mu})Z  \in \hat{\mathcal{H}}.
\end{gather*}
This is a~unique realization in $\hat{\mathcal H}$ (see~\cite{mks07}).
Using $\bigtriangleup P_{\mu}$~\eqref{coproduct}, $\bigtriangleup \hat{x}_{\mu}$~\eqref{c1},
$\bigtriangleup Z=Z\otimes Z$ and relations $\hat{R}_{\mu}$~\eqref{rmu} we obtain coproduct
$\bigtriangleup M_{\mu\nu}$ as in Hopf algebra~\eqref{coproduct}.
Note that the result for~$\bigtriangleup M_{\mu\nu}$ is unique in the~$\kappa$-Poincar\'e--Hopf algebra
$\mathcal{U}_{\kappa}(\mathcal{P})$ since
$\mathcal{U}_{\kappa}(\mathcal{P})\otimes\mathcal{U}_{\kappa}(\mathcal{P})\cap\hat{\mathfrak{K}}=0$ (which is obvious).
Similarly we f\/ind $S(M_{\mu\nu})$ within Hopf algebroid which coincides with $S(M_{\mu\nu})$ in Hopf
algebra~\eqref{antipod} (for details see~\cite{jms13b, mss12}).

There is a~question whether $\bigtriangleup P_{\mu}$ and $\bigtriangleup M_{\mu\nu}$ could be
obtained from twist $\mathcal{F}$ expressed in terms of Poincar\'e generators only.
\begin{enumerate}\itemsep=0pt
\item For $a_{\mu}$ light-like, $a^2=0$, such cocycle twist within Hopf algebra approach exists~\cite{remarks}
\begin{gather}
\label{twist}
\mathcal{F}=\exp \left\{a^{\alpha}P^{\beta}\frac{\ln  [1+(aP) ]}{(a P)}\otimes
M_{\alpha\beta}\right\}.
\end{gather}
The cocycle condition for twist $\mathcal{F}$~\eqref{twist} can be checked using the results by Kulish et al.~\cite{kulish}
in the Hopf algebra setting (see also~\cite{anna2013}).

\item For $a_{\mu}$ time- and space-like such twist does not exist within Hopf algebra.
Namely, starting from $\bigtriangleup
P_{\mu}=\mathcal{F}\bigtriangleup_{0}P_{\mu}\mathcal{F}^{-1}$ and $\bigtriangleup
M_{\mu\nu}=\mathcal{F}\bigtriangleup_{0}M_{\mu\nu}\mathcal{F}^{-1}$ one can construct an operator
$\mathcal{F}=\text{e}^f$, where $f=f_{1}+f_{2}+\cdots$ is expanded in $a_{\mu}$ and expressed in terms of Poincar\'e
generators and dilatation only.
In the f\/irst order we found that the result is not unique, namely we have a~one parameter family of solutions
\begin{gather*}
f_1=a^\alpha P^\beta\otimes M_{\alpha\beta}+u \big(M_{\alpha\beta}\otimes a^\alpha P^\beta-a^\alpha P^\beta\otimes
M_{\alpha\beta}-D\otimes (a P)+(a P)\otimes D\big),
\end{gather*}
where $u\in\mathbb{R}$ is a~free parameter.
However there is one solution ($u=0$) that can be expressed in terms of Poincar\'e generators only.
Also up to f\/irst order in $a_{\mu}$ cocycle condition is satisf\/ied and one obtains the correct classical~$r$-matrix (see
equation~(65) in~\cite{Govindarajan-2}).
In the second order for $f_{2}$ we found a~two parameter family of solutions.
Here there is no solution without including dilatation, that is the operator $\mathcal{F}$ can not be expressed in terms
of Poincar\'e generators only.
We have checked that the corresponding quantum $R$-matrix obtained using $f_{1}$ and $f_{2}$ is correct up to the second
order.
The cocycle condition is no longer satisf\/ied in the Hopf algebra approach, that is $\mathcal{F}$ is not a~twist in the
Drinfeld sense.
However, after using tensor exchange identities~\cite{jms13b, jms13a, mss12} the cocycle condition is sa\-tis\-f\/ied and
$\mathcal{F}$ is a~twist in Hopf algebroid approach.
It also reproduces the~$\kappa$-Poincar\'e--Hopf algebra (when applied to Poincar\'e generators) (see~\cite{jms13a}).
In~\cite{jms13a}, we have developed a~ge\-ne\-ral method for calculating operator $\mathcal{F}$ for a~given coproducts of~$x_{\mu}$ and~$p_{\mu}$.
In Section~3 of~\cite{jms13a} the operator $\mathcal{F}$ is constructed up to the third order for \textit{natural}
realization (classical basis) and it is shown that this operator $\mathcal{F}$ gives the correct coproduct for
$M_{\mu\nu}$ (see equation~(59) in~\cite{jms13a}) and~$R$-matrix (see equation~(61) in~\cite{jms13a}).
We also stated that this operator $\mathcal{F}$ can not be expressed in terms of~$\kappa$-Poincar\'e generators only
(see~\cite[p.~16]{jms13a}).
From the results for $f_{1}$, $f_{2}$ and $f_{3}$ (see equations~(42), (46), (49) in~\cite{jms13a}) one can show that they could be
rewritten in terms of Poincar\'e generators and dilatation only (after using tensor exchange identities).
For alternative arguments on nonexistence of cocycle twist for~$\kappa$-Poincar\'e--Hopf algebra see~\cite{luk, bp10}.
\end{enumerate}

The main point that we want to emphasize is that the twist operator exists within Hopf algebroid approach, that the
cocycle condition is satisf\/ied~\cite{ jms13b,jms13a, mss12} and that this twist gives the full~$\kappa$-Poincar\'e--Hopf
algebra (when applied to the generators of Poincar\'e algebra).

General statements on associativity of star product, twist and cocycle condition in Hopf algebroid approach are:
\begin{enumerate}\itemsep=0pt
\item Lorentz generators $M_{\mu\nu}$ can be written in terms of $x^{(\lambda)}$ and
$p^{(\lambda)}$~\eqref{e13}--\eqref{e16}.
This def\/ines the family of basis labeled by~$\lambda$.
The momenta $p^{(\lambda)}$ do not transform as vectors under~$M_{\mu\nu}$.
The star product is associative for all $\lambda \in \mathbb{R}$.
The corresponding twist $\mathcal{F}^{(\lambda)}$ given in~\eqref{e1} is Abelian and satisf\/ies the cocycle condition for
all $\lambda\in\mathbb{R}$.
Applying $\mathcal{F}^{(\lambda)}$ to primitive coproduct $\bigtriangleup_{0}M_{\mu\nu}$ leads
to~$\kappa$-deformed $\mathfrak{igl}(n)$ Hopf algebra (see~\cite{byk2009,Govindarajan-2, jms2013,klry09, kmps12}).
However, if we apply conjugation by $\mathcal{F}^{(\lambda)}$ to
$\bigtriangleup^{(\lambda)}_{0}M^{(\lambda)}_{\mu\nu}$ (which is not primitive coproduct) we obtain, in the
Hopf algebroid approach~\cite{mss12}, the correct coproduct
$\bigtriangleup^{(\lambda)}M^{(\lambda)}_{\mu\nu}$ corresponding to~$\lambda$ basis (for $\lambda=0$
see~\cite{jms13b}).
Similarly for the antipode~$S$.
Hence, the~$\kappa$-Poincar\'e--Hopf algebra can be obtained by twist $\mathcal{F}$ in the more generalized sense,
i.e.~in the Hopf algebroid approach.
\item If star product is associative in one base, then it is associative in any other base obtained by similarity
transformations~\cite{jms13a}.
\item If star product is associative, then the corresponding twist $\mathcal{F}$ satisf\/ies cocycle condition in the Hopf
algebroid approach, and vice versa.
Note that, there exist star products which are associative but the corresponding twist operator $\mathcal{F}$ does not
satisfy the cocycle condition in the Hopf algebra approach.
\end{enumerate}

\section{Final remarks}

It is important to note that the work presented in this paper is not genuinely dif\/ferent from Lu's construction of Hopf
algebroid~\cite{l95} and that we use a~particular choice of the algebra which makes it easier to construct the coproduct
as an algebra homomorphism to the subalgebra ${\hat{\mathcal{B}}}/{\hat{\mathfrak{I}}}$.
By this particular choice of algebra we are able to satisfy
\begin{gather*}
m(1\otimes{S})\bigtriangleup={\hat{\alpha}}{\hat{\epsilon}},
\qquad
m({S}\otimes 1)\bigtriangleup={\hat{\beta}}{\hat{\epsilon}}{S},
\end{gather*}
while in~\cite{l95} $m(1\otimes{S})\bigtriangleup$ is not well def\/ined (for the version of
coproduct in~\cite{l95}) because $m(1\otimes {S}){\hat{\mathfrak{K}}}\neq0$, while in our case
$m(1\otimes{S}){\hat{\mathfrak{I}}}=0$.
Therefore we do not need the section~$\gamma$ in the f\/irst identity for the antipode.
In our approach, since we have
$\bigtriangleup:{\hat{\mathcal{H}}}\rightarrow{\hat{\mathcal{B}}}/{\hat{\mathfrak{I}}}
=\bigtriangleup{\hat{\mathcal{H}}}$
and
\begin{gather*}
m(1\otimes{S}){\hat{\mathfrak{I}}}=0,
\end{gather*}
it is easy to see that~\eqref{nova} holds $\forall\, h\in{\hat{\mathcal{H}}}$.
We are doing this in order to explain the structure of quantum phase space, i.e.~Weyl algebra
${\hat{\mathcal{H}}}$.

An axiomatic treatment of the Hopf algebroid structure on general Lie algebra type noncommutative phase spaces,
involving completed tensor products, has recently been proposed in~\cite{skoda}.

The construction of QFT suitable for~$\kappa$-Minkowski spacetime is still under active research~\cite{klm00-,klm00,
ms11}.
We plan to apply~$\kappa$-deformed phase space, Hopf algebroid approach and twisting to NCQFT and NC (quantum) gravity.

\subsection*{Acknowledgements}

The authors would like to thank A.~Borowiec, J.~Lukierski, A.~Pachol, R.~\v{S}trajn and Z.~\v{S}koda for useful
discussions and comments.
The authors would also like to thank the anonymous referee for useful comments and suggestions.

\pdfbookmark[1]{References}{ref}
\LastPageEnding

\end{document}